\documentclass[runningheads]{llncs}
\usepackage[T1]{fontenc}
\usepackage{graphicx}
\usepackage{amsmath}
\usepackage{amssymb}
\usepackage{booktabs}
\usepackage{placeins}
\usepackage{algorithm}
\usepackage{algorithmic}
\usepackage{xcolor}
\usepackage{todonotes}
\usepackage{changes}
\usepackage{bbding}

\begin{document}

\title{Active Inference-Based Adaptive Routing for Heterogeneous Edge AI Services}
\titlerunning{AIF-Based Adaptive Routing for Edge AI}
\authorrunning{Wang et al.}

\author{
  Zihang Wang\inst{1} \Envelope \and
  Boris Sedlak\inst{2} \and
  Schahram Dustdar\inst{1,2}
}

\institute{
  Distributed Systems Group, TU Wien, Vienna, Austria\\
  \email{\{zihang.wang, dustdar\}@dsg.tuwien.ac.at}
  \and
  Universitat Pompeu Fabra, Barcelona, Spain\\
  \email{\{boris.sedlak, schahram.dustdar\}@upf.edu}
}

\maketitle

\begin{abstract}
Edge computing enables AI inference closer to data sources, reducing latency and bandwidth costs.
However, orchestrating AI services across the cloud-edge continuum remains challenging due to dynamic workloads and infrastructure variability. 
We present \textbf{AIF-Router}, an Active Inference--based routing framework that autonomously learns to balance latency, throughput, and resource utilization across multi-tier AI services without offline training. 
AIF-Router performs Bayesian state inference and expected free energy minimization to guide routing decisions based on observability-driven real-time metrics. 
 
Despite device instability on edge nodes, AIF-Router exhibits stable online learning behavior and demonstrates the feasibility of applying Active Inference for adaptive AI service orchestration in unreliable edge environments. 
Our findings highlight both the promise and practical challenges of deploying self-adaptive decision-making frameworks for real-world edge AI systems.
\end{abstract}

\section{Introduction}
\label{sec:introduction}

Edge AI systems deploy inference services across the cloud-edge continuum---from resource-constrained devices (e.g., NVIDIA Jetson) to cloud servers---to meet latency and QoS requirements~\cite{zhou2019edge,deng2020edge}.
Routing inference requests across such multi-tier deployments requires balancing latency, resource cost, and reliability under dynamic workloads~\cite{chen2012empirical,zhao2017tasks}.
However, achieving adaptive routing in edge environments faces three fundamental challenges: (1)~\textit{partial observability}---backend resource states must be inferred from aggregated metrics; (2)~\textit{infrastructure instability}---edge devices experience pod restarts and service disruptions; (3)~\textit{online adaptation}---systems must learn from runtime feedback without environment-specific pretraining.

Existing approaches cannot address all three challenges simultaneously. Static policies (e.g., round-robin) fail to adapt to workload shifts and resource heterogeneity~\cite{burns2016borg}.
Reinforcement learning-based schedulers~\cite{mao2019decima,xiao2018gandiva} achieve adaptive routing but require costly offline training and struggle with distribution shift when deployed environments differ from training conditions.
Control-theoretic methods~\cite{hellerstein2004feedback} offer theoretical guarantees but depend on accurate system models that are difficult to obtain in heterogeneous edge settings.
However, existing work lacks a principled framework for online adaptive routing that learns directly from runtime observations without requiring prior system knowledge or offline training.

\begin{figure*}[!htp]
\centering
\includegraphics[width=0.95\textwidth]{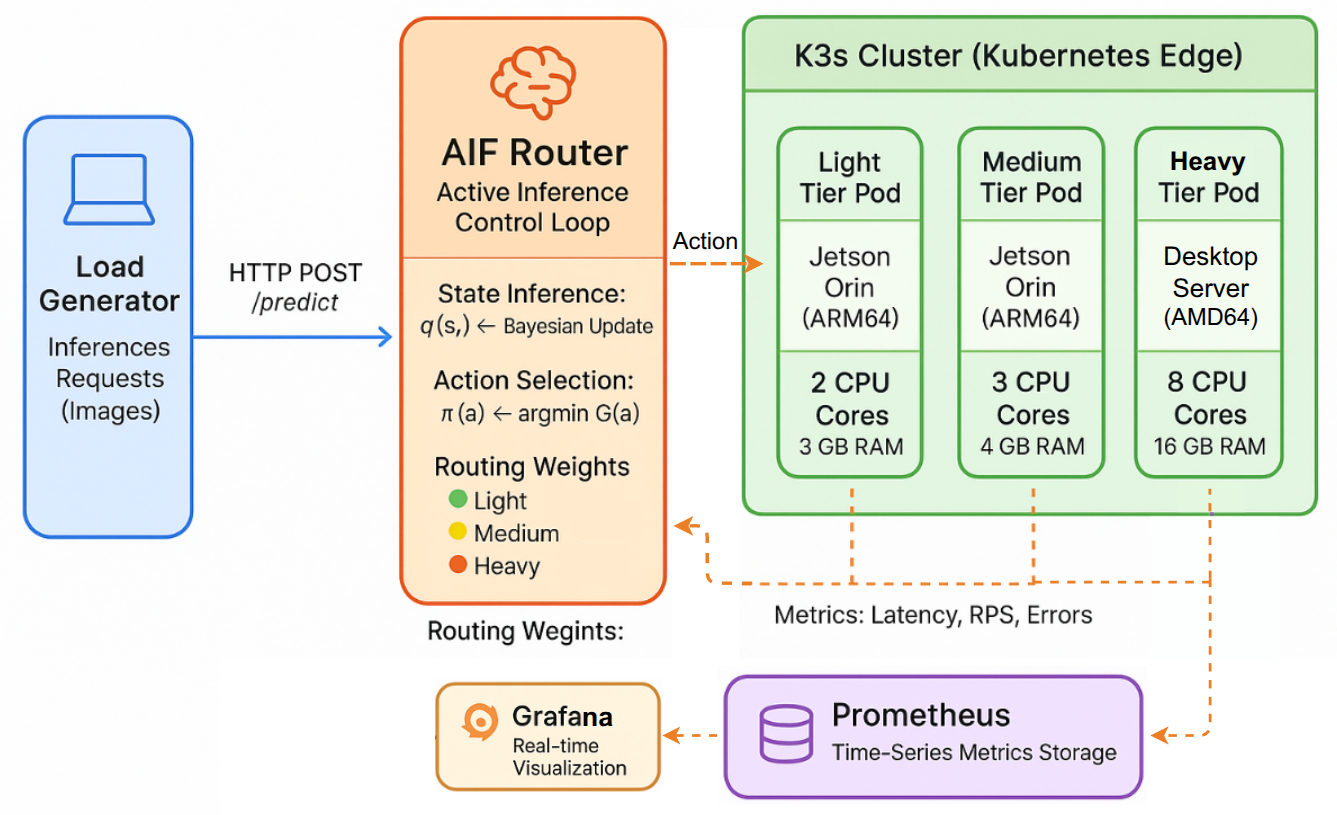}
\caption{AIF-Router control flow with Bayesian state inference, action selection, and multi-tier request dispatching.}
\label{fig:architecture}
\end{figure*}

This paper investigates whether Active Inference~\cite{friston2017active,parr2022active}---a Bayesian framework for online learning and decision-making under uncertainty -- can enable adaptive edge AI routing without offline training.
We present \textbf{AIF-Router}, an Active Inference framework that maintains probabilistic beliefs over system states and selects routing actions by minimizing expected free energy, enabling zero-shot deployment without offline training.
Figure~\ref{fig:architecture} illustrates the AIF-Router control flow.

As shown in Figure~\ref{fig:architecture}, the router performs Bayesian state inference from observed metrics (latency, throughput, errors), computes expected free energy for candidate actions, and continuously learns the observation model $A$ and the transition model $B$ online. The system orchestrates requests across three heterogeneous tiers (Light/Medium/Heavy) deployed on a K3s cluster, balancing goal achievement with exploration.

\textbf{Contributions.}
This paper makes the following contributions:
(1)~We demonstrate that Active Inference enables adaptive edge AI routing without offline training, reducing deployment barriers and enabling immediate adaptation to diverse edge environments.
(2)~We show that Bayesian online learning can maintain stable routing behavior despite infrastructure instability, providing an approach for orchestration in unreliable edge environments where devices experience service disruptions.
(3)~We identify a latency-reliability tradeoff in adaptive edge orchestration, revealing that adaptive routing can improve performance but may impact reliability, offering insights for designing robust orchestration systems.

\paragraph{Paper Organization.}
The remainder is organized as follows: Section~\ref{sec:related} reviews related work; Section~\ref{sec:architecture} describes the system architecture; Section~\ref{sec:methodology} presents the Active Inference methodology; Section~\ref{sec:experiments} reports the experimental results; and Section~\ref{sec:conclusion} concludes.

\section{Related Work}
\label{sec:related}

\textbf{Edge AI orchestration.}
Early work in edge computing studied where to run inference, focusing on model partitioning and offloading between edge and cloud~\cite{kang2017neurosurgeon}. Later systems such as Clipper~\cite{crankshaw2017clipper}, Clockwork~\cite{gujarati2020clockwork}, and specialized serving frameworks such as INFaaS~\cite{romero2021infaas} and InferLine~\cite{crankshaw2020inferline} improved serving efficiency through batching, GPU sharing, and model-level optimization. These approaches, however, rely on static or heuristic routing and adapt poorly to runtime dynamics.
Federated and collaborative inference
approaches~\cite{mcmahan2017communication,teerapittayanon2017distributed} distribute
computation across devices but focus primarily on training or fixed inference
pipelines. Self-adaptive microservice
routing~\cite{zhao2019aiops,pahl2018microservices} addresses orchestration in
cloud-native environments but typically assumes stable infrastructure and homogeneous
resources, unlike edge scenarios with device failures and heterogeneous capacities.

\textbf{Adaptive load balancing.}
Reinforcement learning has been applied to scheduling and resource management~\cite{mao2016resource,tesauro2006hybrid,mao2019decima,xiao2018gandiva}, but usually requires long offline training and retraining when workloads shift. Recent systems like Decima~\cite{mao2019decima} learn scheduling algorithms via RL for data processing clusters, while Gandiva~\cite{xiao2018gandiva} optimizes GPU cluster scheduling for deep learning workloads. Control-based methods~\cite{hellerstein2004feedback} offer stability guarantees but depend on accurate system models that are hard to obtain in heterogeneous edge settings. Bayesian methods~\cite{snoek2012practical} and multi-armed bandit algorithms~\cite{auer2002finite} provide principled approaches to online learning under uncertainty, but have seen limited application to service orchestration at scale.
Lightweight reinforcement learning approaches such as Thompson
Sampling~\cite{russo2018tutorial} and contextual bandits~\cite{auer2002finite} offer
faster convergence than deep RL, but still require explicit reward engineering and
struggle with delayed feedback in complex orchestration scenarios. Bayesian
optimization methods~\cite{frazier2018tutorial,snoek2012practical} excel at
hyperparameter tuning but scale poorly to high-dimensional action spaces and
continuous decision-making. Active Inference differs by maintaining explicit
probabilistic beliefs over system states, learning observation and transition models
online from raw metrics, and unifying exploration-exploitation through expected free
energy minimization without requiring hand-crafted reward functions.

\textbf{Active inference in systems.}
Active Inference (AIF) provides a Bayesian perspective on adaptive control and learning~\cite{friston2017active,parr2022active}. It has been explored in robotics and autonomous control~\cite{lanillos2020active}, where systems learn directly from sensory feedback without predefined rewards.
Transferring this to service computing, Sedlak et al.~\cite{sedlak_multi-dimensional_2025,sedlak_adaptive_2024} used AIF to improve autoscaling through robust generative models.
However, to the best of our knowledge, AIF has not yet been applied to service routing or edge AI orchestration. Our work takes this step, using AIF as a lightweight, online decision framework for adaptive inference scheduling.

\paragraph{Research Gap.}
Existing approaches achieve adaptive scheduling through offline training (RL-based systems~\cite{mao2019decima,xiao2018gandiva}), accurate system models (control-theoretic methods~\cite{hellerstein2004feedback}), or static heuristics~\cite{kang2017neurosurgeon} that cannot learn from runtime feedback. Edge AI routing demands online learning under partial observability, infrastructure instability, and multi-objective optimization (latency, throughput, reliability) without offline training. Active Inference addresses this gap through Bayesian online learning and principled exploration-exploitation via expected free energy minimization.

\section{System Architecture}
\label{sec:architecture}

As illustrated in Figure~\ref{fig:architecture}, AIF-Router orchestrates inference requests across three heterogeneous tiers: a \textbf{light tier} (2 CPU cores on Jetson Orin), a \textbf{medium tier} (3 CPU cores on Jetson Orin), and a \textbf{heavy tier} (8 CPU cores on desktop server).
Each tier runs a ResNet-50 ONNX model exposing a unified HTTP inference endpoint.
The router collects request-level metrics (latency, throughput, error rate) at 1-second intervals to drive Bayesian belief updates, and queries aggregated resource metrics (per-tier CPU utilization) every 10 seconds to enrich state representation.
This heterogeneous deployment---where edge devices have constrained resources and experience service disruptions, while cloud servers offer higher capacity---motivates the state/action/observation design in Section~\ref{sec:methodology}.

The system is deployed on a K3s cluster spanning two NVIDIA Jetson Orin devices (edge) and one AMD Ryzen desktop server (cloud), providing a realistic testbed for evaluating online adaptive routing under infrastructure heterogeneity and instability.

\section{Active Inference Routing Methodology}
\label{sec:methodology}

AIF-Router applies Active Inference~\cite{friston2017active} to online routing: it maintains probabilistic beliefs over system states, learns observation and transition models from experience, and selects actions by minimizing expected free energy---balancing immediate performance goals against information-gathering exploration.

\subsection{State, Action, and Observation Design}
\label{sec:spaces}

\paragraph{State Space.}
We model the system state $s_t$ as a 5-dimensional discrete grid encoding both global load conditions
and per-tier resource availability:
\[
s_t = (\ell_t, r_t, u^H_t, u^M_t, u^L_t) \in \{0,1,2\}^5, \quad |\mathcal{S}| = 243
\]
where $\ell_t$ represents latency level (low/medium/high), $r_t$ request rate level,
and $u^H_t, u^M_t, u^L_t$ per-tier CPU utilization (idle/moderate/saturated).
The resulting 243-state space enables real-time Bayesian inference suitable for online control.

\paragraph{Action Space.}
An action $a$ specifies routing weights $(w_L, w_M, w_H)$ over the three tiers.
Rather than continuous optimization over the weight simplex, we predefine 20 discrete policies:
\begin{itemize}
    \item 1 balanced policy: $(0.33, 0.33, 0.34)$
    \item 5 heavy-biased policies: $(0.15, 0.25, 0.60)$ to $(0.0, 0.0, 1.0)$
    \item 4 medium-biased, 4 light-biased, and 6 adaptive/exploratory policies
\end{itemize}

Discrete actions simplify the planning problem by reducing expected free energy computation to evaluation over a finite candidate set, while maintaining interpretability---operators can inspect learned routing strategies.
The chosen set spans from uniform load balancing to extreme concentration, ensuring sufficient coverage of the strategy space.
\paragraph{Observation Space.}
Every second, the router observes $o_t = (\hat{\ell}_t, \hat{r}_t, \hat{q}_t, \hat{e}_t)$:
P95 latency, request rate, queue depth, and error rate (discretized into 2--3 bins).
These aggregated metrics cannot directly reveal per-tier backend states ($u^H_t, u^M_t, u^L_t$), which must be inferred via Bayesian belief updates.

\subsection{Generative Model and Adaptive Preferences}
\label{sec:generative}

AIF-Router maintains three learned models that together define its understanding of the environment.

\paragraph{Observation Model $A$: Linking States to Observations.}
The observation model $p(o_t | s_t)$ specifies how likely each observation is under each state.
We factorize this into four independent components (latency, RPS, queue, error),
each represented as a matrix mapping 243 states to 2--3 observation bins:
\[
p(o_t | s_t) = p(\hat{\ell}_t | s_t) \cdot p(\hat{r}_t | s_t) \cdot p(\hat{q}_t | s_t) \cdot p(\hat{e}_t | s_t)
\]
Initially, these matrices are uniform (reflecting no prior knowledge).
As the router observes transitions, it updates $A$ matrices via Bayesian pseudo-count accumulation (Section~\ref{sec:learning}).

\paragraph{Transition Model $B$: Predicting Action Outcomes.}
The transition model $p(s_{t+1} | s_t, a)$ encodes how actions affect system state.
For each of the 20 actions, we maintain a separate $243 \times 243$ transition matrix.
This allows the router to learn, for example, that "routing 80\% to heavy tier" reliably reduces latency
when starting from high-load states, but may fail under edge device instability.

Unlike simulation-based methods that assume known dynamics, edge systems exhibit hardware variability (Jetson restarts, thermal throttling), workload non-stationarity (burst vs.\ steady traffic), and deployment-\allowbreak{}specific characteristics (network latency between nodes).
Learning $B$ online from experience enables zero-shot deployment across diverse environments without requiring prior system knowledge or offline model training.

\paragraph{Preference Distribution $C$: Encoding Goals.}
Active Inference agents encode desired outcomes via a preference distribution $C(o)$ over observations.
We define (in log-probability space):
\[
C(o) = C_{\ell}(\hat{\ell}) + C_{r}(\hat{r}) + C_{q}(\hat{q}) + C_{e}(\hat{e})
\]
where each component assigns higher preference to favorable observations
(e.g., $C_{\ell}$ strongly prefers low-latency bins, $C_{e}$ strongly prefers low error rates).

\textbf{Adaptive preference adjustment.}
A unique challenge in unreliable edge environments is infrastructure instability
(e.g., Jetson nodes frequently restarting under load).
Static preferences that always prioritize latency can lead to aggressive routing toward unstable tiers,
amplifying failure rates.

To address this, we introduce \emph{dynamic preference adjustment}:
the router monitors recent error rates and automatically modulates $C_{\ell}$ and $C_{e}$.
When error rate exceeds 15\%, the router relaxes latency preferences and dramatically increases
error avoidance (shifting $C_{e}$ from $-3.0$ to $-11.5$ in log space).
This allows the system to prioritize reliability during instability while optimizing performance under normal conditions.

This adaptive mechanism is critical to our experimental results (Section~\ref{sec:experiments}):
without it, the router aggressively routes to unstable edge tiers, achieving low latency but with significantly elevated failure rates.

\paragraph{Integration: Inference-Action-Learning Cycle.}
The three matrices converge to form a closed adaptive control loop:
\begin{center}
\small
$\boxed{o_t} \xrightarrow{\mathbf{A}} \boxed{q(s_t)}
\xrightarrow{\mathbf{B},\mathbf{C}} \boxed{G(a)}
\xrightarrow{\arg\min} \boxed{a_t}
\xrightarrow{\text{execute}} \boxed{o_{t+1}}$
\\[0.3em]
\hspace{2em}$\circlearrowleft$ \textit{Fast loop (1s): Bayesian inference and action selection} \hspace{2em}
\\[0.5em]
$\boxed{\text{experience}} \xrightarrow{\text{batch update}}
\boxed{\mathbf{A}, \mathbf{B} \text{ updated}}$
\\[0.3em]
\hspace{2em}$\circlearrowleft$ \textit{Slow loop (10s): Model learning} \hspace{2em}
\end{center}
At each step, matrix $\mathbf{A}$ inverts observed metrics to posterior beliefs;
$\mathbf{B}$ predicts state trajectories under candidate actions; $\mathbf{C}$ evaluates
trajectories against preferences; and expected free energy---as in Eq.~\eqref{eq:efe}---
integrates these components to select actions that balance goal achievement and exploration.
Simultaneously, accumulated observations refine $\mathbf{A}$ and $\mathbf{B}$, closing the
adaptation loop.
\subsection{Action Selection via Expected Free Energy}
\label{sec:efe}

Given belief $q(s_t)$, the router selects actions by minimizing expected free energy:
\begin{equation}
G(a) = \underbrace{\text{Risk}(a)}_{\text{goal achievement}} + \underbrace{\text{Ambiguity}(a)}_{\text{exploration}} + \underbrace{\text{Cost}(a)}_{\text{regularization}}
\label{eq:efe}
\end{equation}
where Risk measures expected divergence from preferred observations $C$, Ambiguity encourages exploration of uncertain states, and Cost penalizes extreme routing policies. Actions are sampled via softmax: $p(a) \propto \exp(-\beta \cdot G(a))$ with $\beta = 5.0$.

\subsection{Online Learning with Timescale Separation}
\label{sec:learning}

AIF-Router refines its generative model ($A$, $B$) through online learning,
decoupling fast inference from slow model updates to maintain stability.

\paragraph{Fast Loop: Bayesian Belief Update (1 second).}
Every second, the router updates its belief via:
\begin{equation}
q(s_t | o_{1:t}) \propto p(o_t | s_t) \cdot p(s_t | o_{1:t-1})
\label{eq:belief}
\end{equation}
where the prior $p(s_t | o_{1:t-1}) = B_{a_{t-1}} \cdot q(s_{t-1})$ is predicted using the transition model,
and the likelihood $p(o_t | s_t)$ comes from the observation model.
This Bayesian update distributes probability mass over the 243 states from observed evidence,
inferring hidden backend loads from aggregated metrics.

\paragraph{Slow Loop: Model Learning (10 seconds).}
Every 10 seconds, the router performs batch updates to $A$ and $B$ matrices using accumulated observations.

\textbf{Observation model learning ($A$):}
For each observed $(o_t, q(s_t))$ pair, we update pseudo-counts:
\[
A[o_t, :] \leftarrow A[o_t, :] + \alpha \cdot q(s_t), \quad \alpha = 0.05
\]
distributing the update across likely states weighted by the posterior belief.

\textbf{Transition model learning ($B$) with experience replay:}
A key challenge is that transitions immediately after action changes are unreliable---the system has not yet stabilized.
We address this using \emph{sigmoid-weighted learning}:
transitions occurring $\Delta t$ seconds after an action change receive weight
$w(\Delta t) = 1 / (1 + e^{-(\Delta t - 2)/2})$,
smoothly up-weighting observations as the system settles.
This avoids corrupting $B$ matrices with transient dynamics.

We maintain a replay buffer of 5000 recent transitions and sample batches of 100 for each update,
improving sample efficiency and stability.

Timescale separation (1s inference, 10s learning) is critical for stability,
ensuring the router operates on a quasi-static model between updates.

\section{Experimental Evaluation}
\label{sec:experiments}

\subsection{Experimental Setup}

\textbf{Workload.} We use the Tiny-ImageNet-200 test set (100 JPEG images) with burst traffic at 50 RPS. Each strategy
(AIF-Router and Baseline) was tested in 3 repeated 45-minute runs with 15-minute cooldown between runs, totaling 412{,}500
requests.

\textbf{Baseline Strategy.}
For comparison, we implement a fixed-weight router with uniform allocation:
\[
w_\text{light}=0.33, \quad
w_\text{medium}=0.33, \quad
w_\text{heavy}=0.34.
\]
This represents capacity-agnostic strategies commonly used in production systems (e.g., Kubernetes
Services~\cite{burns2016borg}, NGINX upstream) where backend capacities are unknown or change dynamically.
While a capacity-aware baseline weighted by CPU limits (e.g., 0.15/0.23/0.62 reflecting the 2:3:8 core ratio) would provide a
stronger comparison, it requires environment-specific tuning and prior knowledge of tier capacities---precisely the deployment
overhead AIF-Router aims to eliminate.
Our research question is whether online learning can autonomously discover and exploit resource heterogeneity from runtime
observations alone.

\textbf{Metrics.} We report success rate (\%), P50/P95 latency (ms), and tier distribution as mean $\pm$ standard deviation over 3 runs.

\FloatBarrier

\subsection{Performance Results}

Table~\ref{tab:results} summarizes AIF-Router
vs.\ baseline over the 3 repeated 45-minute runs
(mean $\pm$ standard deviation).

\begin{table}[!htp]
\centering
\small
\setlength{\tabcolsep}{2pt}
\caption{Overall performance comparison at
50~RPS.}
\label{tab:results}
\begin{tabular}{lcccccc}
\toprule
Strategy & Succ. (\%) & P50 (ms) & P95 (ms) & Heavy (\%) & Med. (\%) & Lt. (\%) \\
\midrule
AIF-Router & 77.9 $\pm$ 0.26 & 2003 $\pm$ 119 & 5318 $\pm$ 11 & 35.8 $\pm$ 0.4 & 30.0 $\pm$ 0.4 & 12.1 $\pm$ 0.4 \\
Baseline   & 89.4 $\pm$ 0.29 & 3067 $\pm$ 33  & 5268 $\pm$ 11 & 34.0 $\pm$ 0.2 & 32.9 $\pm$ 0.3 & 22.4 $\pm$ 0.2 \\
\midrule
\textbf{$\Delta$ (AIF--Base)} & -11.5pp & -35\% & +1\% & +1.8pp & -2.9pp & -10.3pp \\
\bottomrule
\end{tabular}
\end{table}

\begin{figure}[!htp]
\centering
\includegraphics[width=0.55\linewidth]{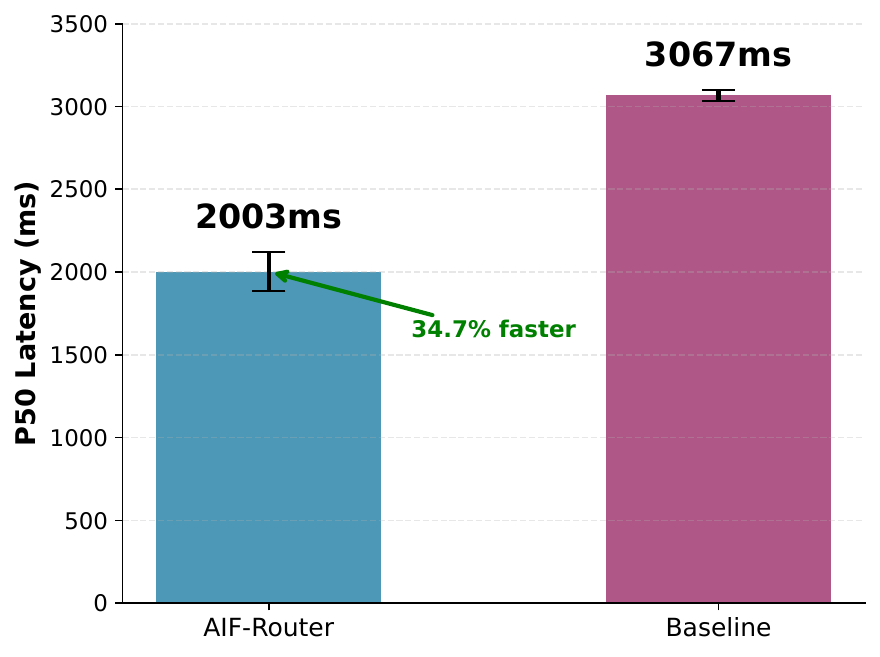}
\caption{\textbf{P50 latency comparison.}
AIF-Router achieves 34.7\% lower median latency (2003~ms vs.~3067~ms, $p<0.0001$).}
\label{fig:latency}
\end{figure}

\FloatBarrier

\begin{figure}[!htp]
\centering
\includegraphics[width=1\linewidth]{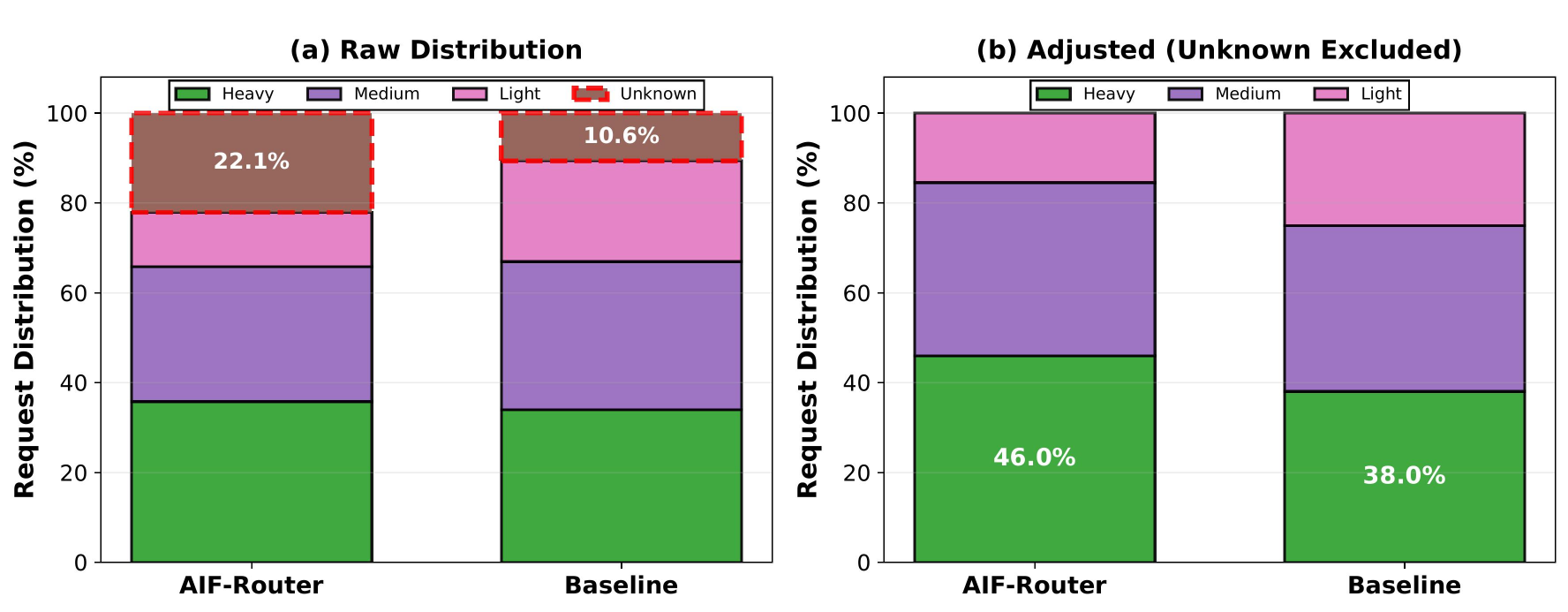}
\caption{\textbf{Tier allocation comparison.}
AIF-Router learns to allocate more requests to the heavy tier (46\% vs 38\%) after observing performance feedback, while experiencing higher failure rates on unstable edge devices.}
\label{fig:tier_distribution}
\end{figure}

\textbf{Key Findings:}
\begin{itemize}
  \item Latency-reliability tradeoff. As illustrated in Figure~\ref{fig:latency}, AIF-Router achieves 34.7\%
lower median latency (2003ms vs 3067ms, $p<0.0001$) but suffers an 11.5 percentage point lower success rate (77.9\% vs
89.4\%, 29.4$\sigma$ statistical significance). This reveals a latency-reliability trade-off in our deployment:
optimizing for low latency can compromise reliability when edge infrastructure is unstable.

  \item Despite lower overall success, AIF-Router exhibits clear adaptive behavior. Among successful requests only,
the tier distribution (Figure~\ref{fig:tier_distribution}b) shows AIF-Router allocates 46\% to heavy tier vs
baseline's 38\%, indicating that Active Inference successfully learns to prefer higher-capacity services in response
to latency observations.

  \item AIF-Router experiences 22.1\% unknown/failed requests compared to baseline's 10.6\%
(Figure~\ref{fig:tier_distribution}a). Investigation reveals that Jetson edge devices (hosting light and medium tiers)
experienced frequent pod restarts during testing (65 restarts for light tier over 4 days), explaining the elevated
failure rate. This highlights a practical challenge: when adaptive routing learns to prefer lower-latency tiers, it
may increase load on already unstable edge devices.

  \item Both strategies show low standard deviation across 3 runs (< 0.3\% for success rate, < 120ms for P50
latency), validating the experimental reproducibility and statistical rigor of our evaluation methodology.
\end{itemize}

\FloatBarrier 

\subsection{Discussion}

\textbf{State Representation Trade-offs.}
AIF-Router employs 3-level discretization (low/medium/high) for each state dimension, resulting in a 243-state space. While
this coarse granularity may hide finer effects, it provides advantages: (1)~noise robustness; (2)~fast Bayesian
inference at 1-second intervals; (3)~improved sample efficiency. This reflects a practical compromise between model
expressiveness and computational tractability for online edge orchestration.

\textbf{Preference Model Design.}
The preference distribution $C(o)$ in AIF-Router is hand-crafted to encode
desired outcomes (low latency, high accuracy, low failure rate). While this approach
leverages domain knowledge to guide initial exploration, it limits adaptability to
evolving operational requirements. Future work should investigate learning the
preference model from historical performance data or administrator feedback, enabling
the router to automatically discover optimal trade-offs between competing objectives
(e.g., cost vs. latency) without manual tuning. Inverse reinforcement learning or
preference learning from rankings could provide principled methods to infer the
underlying objective function from observed expert routing decisions.

\textbf{Limitations.} AIF-Router demonstrates online adaptation but faces challenges:
(1)~higher workloads may require horizontal scaling, as hardware saturation
dominates routing intelligence; (2)~edge device instability (frequent pod
restarts) led to elevated failure rates, suggesting future work should incorporate
service health indicators into the state space. Integrating failure predictors (e.g.,
predictive models of pod restart likelihood based on resource pressure or thermal
conditions) into the observation model would enable proactive avoidance of unstable
tiers before failures occur. (3)~The current cost model does not account for
longer-term action costs such as hardware health degradation or energy consumption.
Learning wear-aware policies that factor in device temperature, battery drain on
mobile edge nodes, or carbon footprint of cloud resources would enable sustainable
orchestration. (4)~Comparison against RL-based schedulers (e.g.,
Decima~\cite{mao2019decima}, Gandiva~\cite{xiao2018gandiva}) and other adaptive
baselines would further clarify Active Inference's trade-offs in sample efficiency
and robustness.

\section{Conclusion}
\label{sec:conclusion}
We presented AIF-Router, an Active Inference–based framework for adaptive routing across heterogeneous edge AI services.
Evaluation with 412K requests showed 34.7\% latency reduction (2003~ms vs.~3067~ms, $p<0.0001$) but 11.5pp lower success rate due to edge device instability---revealing a latency-reliability tradeoff.
Despite infrastructure failures (frequent Jetson pod restarts), AIF-Router autonomously learned to reallocate traffic toward higher-capacity tiers (46\% vs.~38\% heavy allocation) through online Bayesian learning without offline training.
This study demonstrates that Active Inference provides a viable foundation for online decision-making in uncertain edge environments, though future work should integrate service health indicators to improve robustness under infrastructure instability.

\subsection*{Acknowledgement}

This work is supported by the European Union NextGenerationEU/PRTR and CNS2023-144359 financed by MICIU/AEI/10.13039/501100011033. 

\subsection*{Disclosure of Interests}
The authors have no competing interests to declare that are relevant to the content of this
article.

\bibliographystyle{splncs04}
\bibliography{references}

\end{document}